\documentclass[pra, nofootinbib, twocolumn, superscriptaddress]{revtex4-2}
\usepackage{amsmath}
\usepackage{amssymb}
\usepackage{graphics}
\usepackage{epsfig}
\usepackage{verbatim}
\usepackage{textcomp}
\usepackage[colorlinks=true,urlcolor=blue,linkcolor=blue,citecolor=blue]{hyperref}
\usepackage{tabularx}
\usepackage{xcolor}
\usepackage{longtable}
\usepackage{multirow}
\usepackage[normalem]{ulem}

\newcolumntype{C}[1]{>{\centering\arraybackslash}m{#1}}
\newcolumntype{s}[1]{>{\centering\arraybackslash}X}
\makeatletter
\def\cat@comma@active{\catcode`\,12}%
\makeatother

\begin{document}
	
	\title{Tuning the initial phase to control the final state of a driven qubit}
	
	
	\author{Polina~O.~Kofman}
	\email{polinaokofman@gmail.com}
	\affiliation{B.~Verkin Institute for Low Temperature Physics and Engineering, Kharkiv 61103, Ukraine}
	\affiliation{V.~N.~Karazin Kharkiv National University, Kharkiv 61022, Ukraine}
	\affiliation{Theoretical Quantum Physics Laboratory, Cluster for Pioneering Research, RIKEN, Wakoshi, Saitama 351-0198, Japan}
	\affiliation{University of Lisbon and Instituto de Telecomunicações,
Avenida Rovisco Pais 1, Lisboa 1049-001, Portugal}
	\author{Sergey~N.~Shevchenko}
	\affiliation{B.~Verkin Institute for Low Temperature Physics and Engineering, Kharkiv 61103, Ukraine}
	\author{Franco~Nori}
	\affiliation{Theoretical Quantum Physics Laboratory, Cluster for Pioneering Research, RIKEN, Wakoshi, Saitama 351-0198, Japan}
	\affiliation{Quantum Computing Center, RIKEN, Wakoshi, Saitama 351-0198, Japan}
	\affiliation{Department of Physics, The University of Michigan, Ann Arbor, MI 48109-1040, USA}	
	
	\begin{abstract}
		A driven quantum system can experience Landau-Zener-St\"{u}ckelberg-Majorana (LZSM) transitions between its states, when the respective energy levels quasi-cross. If this quasicrossing is traversed repeatedly under periodic driving, the trajectories can interfere either constructively or destructively. In the latter case, known as coherent destruction of tunneling, the transition between the energy states is suppressed. Even for the \textit{double}-passage case, the accumulated phase difference (also referred to as the St\"{u}ckelberg phase) can lead to destructive interference, resulting in no transition. In this paper, we discuss a similar process for the \textit{single}-passage dynamics. We study the LZSM single-passage problem starting from a superposition state. The phase difference of this initial state results in interference. When this results in either a zero or a unit transition probability, such a situation can be called \textit{single-passage complete localization} in a target state. The phase can be chosen so that the occupation probabilities do not change after the transition, which is analogous to the problem of \textit{transitionless driving}. We demonstrate how varying the system parameters (driving velocity, initial phase, initial detuning) can be used for quantum coherent control.
	\end{abstract}
	
	\date{\today }
	\maketitle
	\section{Introduction}
For quantum computing it is important to have different methods on how to steer quantum systems to desired states, see e.g. Refs.~\cite{Emmanouilidou2000, Childress2010, Bason2012, Gagnon2017}. The ability to predict the behavior of the system opens the opportunity to use it either as a quantum logic gate or to improve already existing gates. This means to make the process faster or the errors smaller, or easier for an experimental realization.
 		\begin{figure*}[t]
		\includegraphics[width=2\columnwidth]{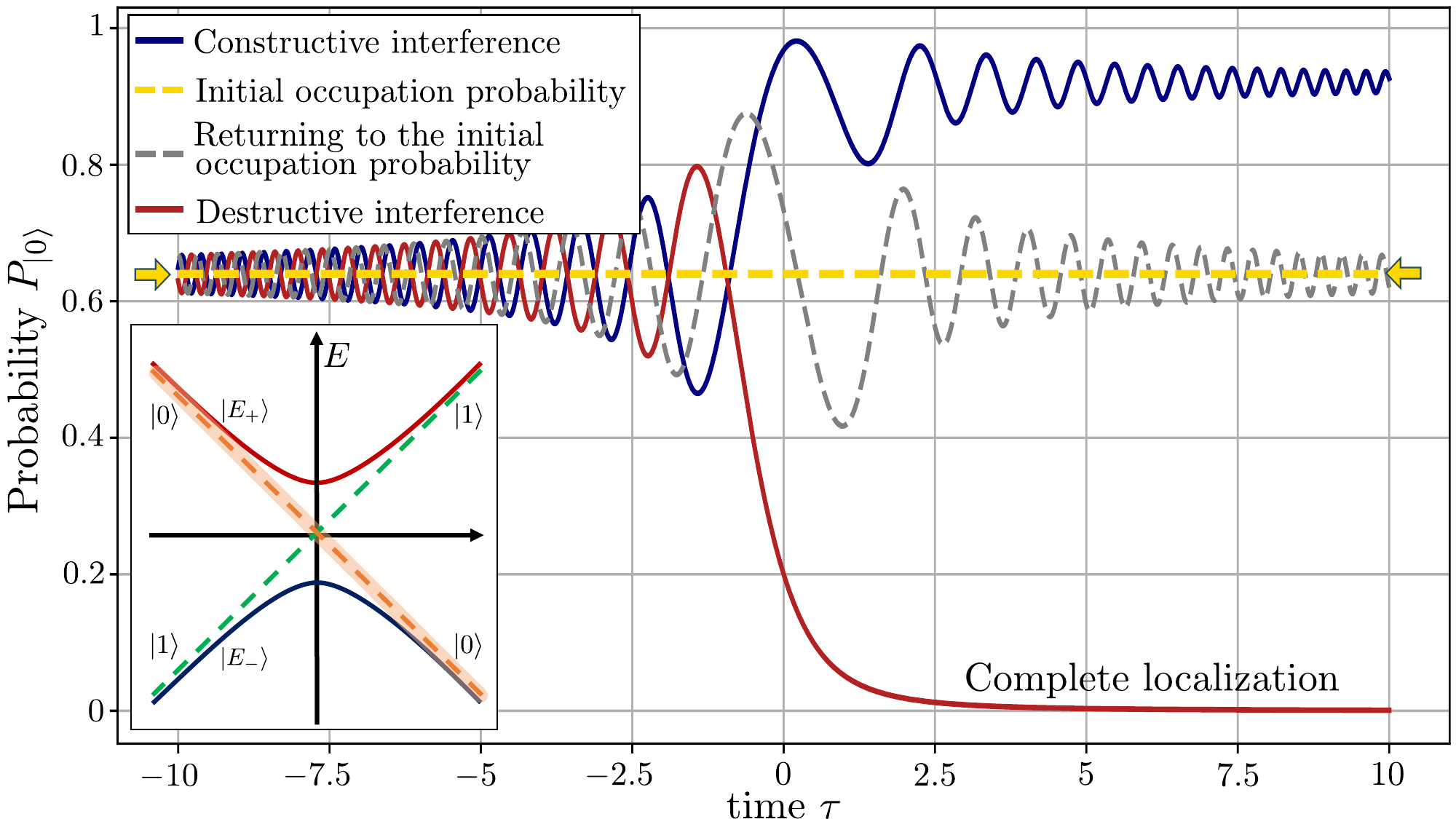}
		\caption{Dependence of the occupation probability on the dimensionless time $\tau=\sqrt{v/(2\hbar)}t$ in the diabatic basis. A graphical illustration of two possible bases is shown in the bottom left inset. The adiabatic basis $\left\vert E_\pm\right\rangle$ refers to the eigenvalues of the energy levels (red and blue hyperbolas). Diabatic basis $\{\left\vert 0\right\rangle, \left\vert 1\right\rangle\}$ is the basis of eigenvalues of the Pauli matrices (orange and green dashed straight lines). The three curves show three possible different occupation probabilities of state $\left\vert 0 \right \rangle$ starting from the same initial occupation probability $P_{\left\vert 0 \right \rangle}$ and \textit{different initial phases} of the wave function. We are interested in the long-time occupation of the ground state to the right (which is $P_{\left\vert 0 \right \rangle}$); for convenience, we say "constructive interference" when this results in increasing $P_{\left\vert 0 \right \rangle}$ and vice versa. So, the maximum value of the occupation probability $P_{\left\vert 0 \right \rangle}$ corresponds to constructive interference (dark blue curve). The phase in this case is defined in Eq.~(\ref{ConstructivePhase}) and approximately equals to $\phi_\mathrm{i}\approx 0.36$. The minimal value of the occupation probability corresponds to destructive interference (dark red curve). The phase, in this case, is defined in Eq.~(\ref{DestructivePhase}) and approximately equals to $\phi_\mathrm{i}\approx 1.53$. Any occupation probability between these values, defined by the constructive and destructive interferences, could be obtained by changing the initial phase. In particular, there are parameters which allow the system to return to its initial state (grey dashed curve). The adiabaticity parameter must correspond to the condition defined in Eq.~(\ref{DeltaST}) and then the initial phase is equal to the value in Eq.~(\ref{PhaseStable}) (in the plot above this approximately equals to $\phi_\mathrm{i}\approx -2.78$).}
		\label{Fig:CDTandRemaining}
	\end{figure*}
	A dynamics that allows the system to return to its \textit{original} state can be useful for quantum information. This can be realized by adiabatically slow driving \cite{Wang2016, Wilson2012, Liu2005, Ashhab2006, Campbell2020}. We could also consider the following question: is it possible to return to the original state using fast processes? In some cases this is possible \cite{Zagoskin2011, Atia2019}. A number of parameters exist that can drive the system. If the signal is periodic, there are special values of the frequency and amplitude which guide the system to its ground state \cite{Grossmann1991a,Grossmann1991b, Grifoni1998, Wubs2010, Miao2016}. Since this results from destructive interference \cite{Grossmann1992, Llorente1992, Kayanuma1994, Kayanuma2008, Ashhab2007, Hu2022}, this is known as coherent destruction of tunneling (CDT) and dynamic localization \cite{Dunlap1986,Kayanuma2008}. 
	
	In a different context, the problem of controlling a quantum system can be formulated as how to correct a given drive so that the system remained in one of the basis states. Such transitionless driving was studied in Ref.~\cite{Berry2009}; first for a generic case, and also for the particular situation of a two-level system with linear drive. This last case corresponds to Landau-Zener-St\"{u}ckelberg-Majorana (LZSM) transitions \cite{Shevchenko2010, Ivakhnenko2023, Glasbrenner2023, Fujikawa1997}. LZSM transitions are interesting in different aspects and processes, including: axion-photon conversion \cite{Carenza2023}, interferometry in a non-Hermitian \textit{N}-body interacting boson system \cite{Wang2023},  interference effects in a qubit \cite{Ono2019, Wen2020, Liul2023, Paila2009, Sillanpaeae2006}, Mach-Zehnder–type interferometry in a superconducting qubit \cite{Oliver2005,Oliver2009}, tunneling under the effect of higher-order dispersion \cite{Cao2023}, spin-flip in the multi-stage Stern–Gerlach experiment \cite{He2023}. LZSM transitions can also be driven with intense laser pulses in the avoided-crossing band structure of graphene's Dirac cones \cite{Heide2021, Heide2020, Heide2019, Higuchi2017, Weitz2023}; the authors of these works demonstrated different transition scenarios, including different transition scenarios were demonstrated, including limiting cases such as CDT and the intermediate case of returning to the initial occupation. 
	
 Here we study similar questions to the ones in Refs.~\cite{Grossmann1991b, Berry2009}: how to steer the system to a given state? Can the system remain in the same state as before the driving (also known as \textit{transitionless driving})? What is needed to direct and guide the system to its ground state or excited state, which we call Complete Localization (CL), in a target state? 
 
 In our case, we consider the simplest linear driving with velocity $v$ and starting from a generic superposition state, using the initial phase difference between the spinor components as a \textit{tunable parameter}. A phase measurement is described in Ref.~\cite{Greenberg2022}. In our approach, considering the phase as a controlling parameter, we follow Ref.~\cite{Wubs2005}.  What we mean is illustrated in Fig.~\ref{Fig:CDTandRemaining}, showing the dynamics of a qubit, when starting from three different superposition states, with the same initial occupation probability but with three different phase differences between the spinor components. The three respective curves show (from bottom to top): complete localization in a $\left\vert 1\right\rangle$ state (constructive interference, bottom curve), remaining at the same position as the initial one (middle curve, transitionless driving), and obtaining maximal occupation probability (constructive interference) shown in the top curve. In general, \textit{changing the initial phase gives the possibility to obtain any probability} between constructive and destructive ones. Just for convenience, studying the long-time occupation $P_{\left\vert 0 \right\rangle}$ of the state $\left\vert 0 \right\rangle$, we call the interference constructive, if its occupation $P_{\left\vert 0 \right\rangle}$ increases; and we call the interference destructive, if the population of the state $\left\vert 0 \right\rangle$ decreases.

	The rest of the paper is organized as follows. In Section~\ref{DAIA} we describe the dynamics and introduce important aspects of the adiabatic-impulse approximation (with details of the adiabatic stage of the dynamics in Appendix~\ref{AppendixA}). In Section~\ref{DOIP} we find the dependence of the final probability on the system's parameters, including the initial phase. In Section~\ref{TWOPPVOP} we analyze the range of the possible values of the final occupation probability. The eventual return of the system to its initial state and single-passage  CL are discussed in Section~\ref{RITISATT}. In Section~\ref{PTSS} we describe how to control the qubit only by changing its phase and linear perturbations. The adiabatic evolution in the diabatic basis is described in Appendix~\ref{AppendixB}. The respective results in the adiabatic basis are presented in Appendix~\ref{AppendixC}. In Appendix~\ref{AppendixD}, we generalize the result for any type of dynamics which could be described by the adiabatic impulse model.
	
	\section{Dynamics: Adiabatic-impulse approximation}
        \label{DAIA}
	Consider the dynamics of a two-level system with a linear perturbation. Such a system is described by the Schr\"{o}dinger equation in the diabatic basis
	\begin{equation}
		i\hbar \frac{\partial }{\partial t}\left\vert \psi \right\rangle =-\frac{1}{2%
		}\left( \Delta \sigma _{x}+\varepsilon(t)\sigma _{z}\right) \left\vert \psi
		\right\rangle ,
  \label{Schrodingerequation}
	\end{equation}%
	where 
 \begin{equation}
     \varepsilon(t)=vt,
 \end{equation}

 	\begin{equation}
		\left\vert\psi\right\rangle= \alpha\left\vert 0\right\rangle+\beta\left\vert 1\right\rangle =
		\begin{pmatrix}
			\alpha \\ 
			\beta
		\end{pmatrix},
        \label{DiabaticWaveFunc}
	\end{equation} 
    $\Delta $ and $v$ are constant values and $\sigma _{i}$ stands for the Pauli matrices.
	Such dynamics describes the Landau-Zener-St\"{u}ckelberg-Majorana (LZSM) transitions, with the excitation probability (see Ref.~\cite{Ivakhnenko2023} and references therein)
	\begin{equation}
		\mathcal{P}=\exp(-2\pi\delta),
	\end{equation}
	(if starting from the ground state), where
 \begin{equation}
     \delta =\Delta ^{2}/ (4v\hbar)
 \end{equation} is the adiabaticity parameter.

	The dynamics with initial time $t_\mathrm{i}$ and final time $t_\mathrm{f}$ can be described by the adiabatic-impulse approximation \cite{Kofman2023}. The dynamics could be separated into three different stages \cite{Ashhab2007}: adiabatic evolution, transition, and adiabatic evolution again:
	\begin{equation}
		\left\vert\psi _{\mathrm{f}}\right\rangle=U_{\mathrm{ad}}(t_{\mathrm{f}}, 0)N U_{\mathrm{ad}}(0,t_{\mathrm{i}})\left\vert\psi _{\mathrm{i}}\right\rangle,
        \label{AIMevolution}
	\end{equation}
	where $\left\vert\psi _{\mathrm{i,f}}\right\rangle$ are the initial and final wave functions, $U_{\mathrm{ad}}$ describes the adiabatic evolution, and $N$ describes the LZSM transition. The adiabatic-evolution matrix before the transition is 
	\begin{equation}
		U_{\mathrm{ad}}(0, t_\mathrm{i})=\left( 
		\begin{array}{cc}
			\exp \left( -i\zeta \right) & 0 \\ 
			0 & \exp \left( i\zeta \right)%
		\end{array}%
		\right),
	\end{equation}%
	and after the transition the adiabatic-evolution matrix is
	\begin{equation}
		U_{\mathrm{ad}}(t_\mathrm{f}, 0)=\left( 
		\begin{array}{cc}
			\exp \left( i\zeta \right) & 0 \\ 
			0 & \exp \left( -i\zeta \right)%
		\end{array}%
		\right) .
	\end{equation}
	Here the phase $\zeta $ and its asymptotic expressions at large times, i.e. at 
 \begin{equation}
     t=\pm \tau _{\mathrm{a}}\sqrt{\frac{2\hbar}{v} }
 \end{equation}
     with $\tau _{\mathrm{a}}\gg1$, are the following \cite{Kofman2023}: 
	\begin{eqnarray}
		\zeta \left( \pm \tau _{\mathrm{a}}\right) &=&\frac{1}{2\hbar }%
		\int\limits_{0}^{\pm \tau _{\mathrm{a}}}\sqrt{\Delta ^{2}+2\hbar v\tau^{2}}%
		d\tau \approx \\
		&\approx &\pm \left[ \frac{\tau _{\mathrm{a}}^{2}}{2}+\frac{\delta }{2}-\frac{\delta }{2}\ln {\delta }+\delta \ln {\sqrt{2}\tau _{\mathrm{a}}}\right]
		.
		\label{Zeta}
	\end{eqnarray}
	
	A non-adiabatic transition is described by the transfer matrix $N$, which is associated with a scattering matrix in scattering theory \cite{Moskalets2011}. The components of the transfer matrix are related to the amplitudes of the respective states of the system in energy space.  The diagonal elements of $N$ \cite{Shevchenko2010, Ivakhnenko2023} correspond to the square root of the reflection coefficient $R$, and the off-diagonal elements correspond to the square root of the transmission coefficient $T$ and its complex conjugate: 
	\begin{equation}
		N=\left( 
		\begin{array}{cc}
			\sqrt{R} & \sqrt{T} \\ 
			-\left (\sqrt{T}\right )^{\ast } & \sqrt{R}%
		\end{array}%
		\right) .
	\end{equation}%
	In our problem, these elements are
	\begin{equation}
		R=\mathcal{P}\text{ \ \ and \ \ }T=(1-\mathcal{P})\exp \left( i2\varphi _{%
			\mathrm{S}}\right) ,
	\end{equation}%
	where $\varphi _{\mathrm{S}}$ is the Stokes phase 
	\begin{equation}
		\varphi _{\mathrm{S}}=\frac{\pi }{4}+\mathrm{Arg} \left[ \Gamma \left( 1-i\delta
		\right) \right] +\delta \left( \ln {\delta }-1\right) ,
		\label{PhiStokes}
	\end{equation}
    where $\Gamma$ denotes the Gamma special function.
    In what follows we will use the formalism summarized in this section (and explained at length in Refs.~\cite{Shevchenko2010,Kofman2023}) to describe the system dynamics when starting from a superposition state. Using these results it is possible to describe not only a single passage but also the multiple-passages case using the adiabatic-impulse model. 

 \section{Dependence of the final state on the initial phase}
         \label{DOIP}
	Consider the dependence of the final occupation probability on the initial phase of the wave function. For this, we take the wave function in the diabatic basis Eq.~(\ref{DiabaticWaveFunc}). 

Then taking the initial wave function $\left\vert\psi_\mathrm{i}\right\rangle$ with the component  $\alpha_\mathrm{i}$ real for the initial condition, we introduce the phase difference in  $\beta_\mathrm{i}$:

	\begin{equation}
		\beta_\mathrm{i}=\sqrt{1-\alpha_\mathrm{i}^2} \exp{(i\phi_\mathrm{i})}.
        \label{InitilState}
	\end{equation}

    \begin{figure}[h!]
		\includegraphics[width=0.84\columnwidth]{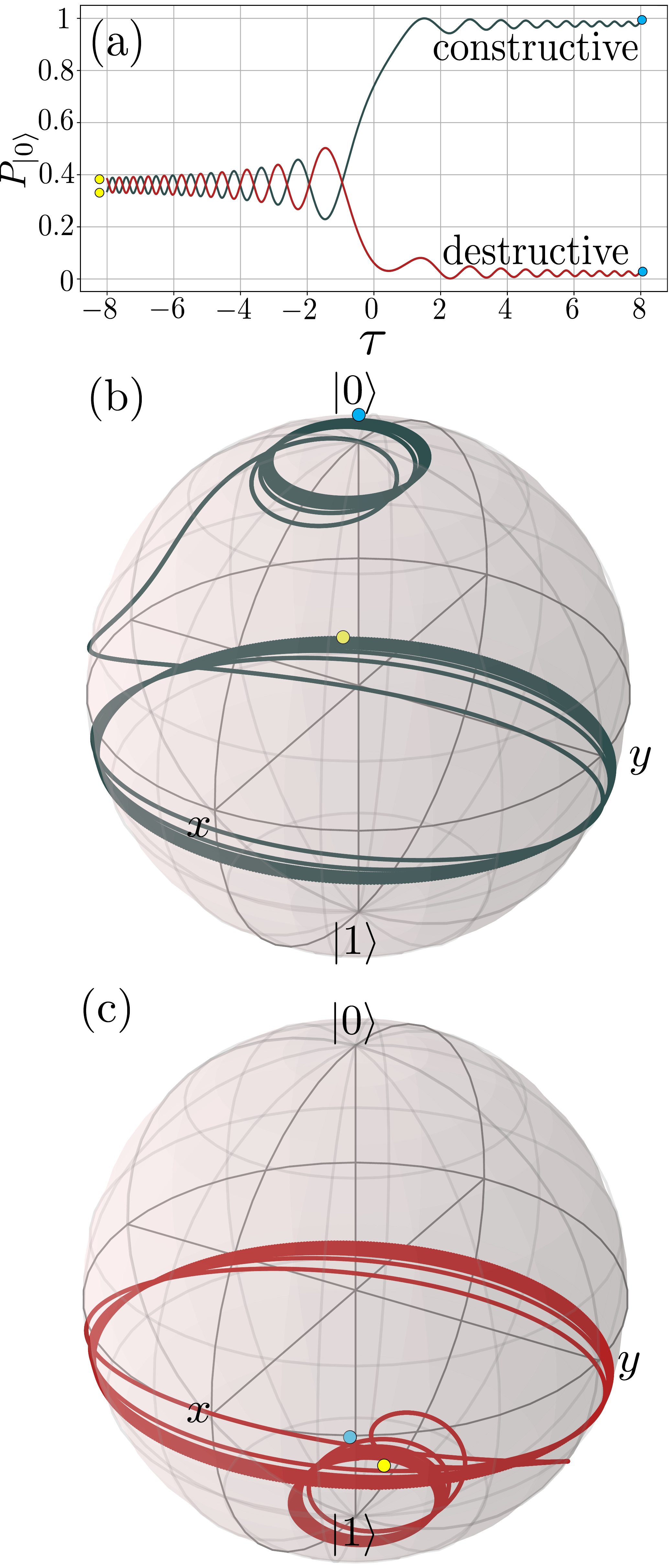}
		\caption{Impact of the initial phase $\phi_\mathrm{i}$ on the final occupation probability $P_{\left\vert0 \right\rangle}$. If the value of the interference term in Eq.~(\ref{AlphaInt}) is maximal, the final occupation probability is also maximal and corresponds to the constructive interference case. On the other hand, if Eq.~(\ref{AlphaInt}) is minimal, the final occupation probability $P_{\left\vert0 \right\rangle}$ is also minimal, and this corresponds to the destructive interference case. (a)~shows the temporal dependence of both of these occupation probabilities. (b)~shows the dynamics on the Bloch sphere for the \textit{con}structive interference case shown in panel~(a). (c)~shows the dynamics on the Bloch sphere for the \textit{de}structive interference case shown in panel~(a). The initial occupation probability is $P_{\left\vert 0 \right\rangle \mathrm{i}}=0.36$ and the adiabaticity parameter is $\delta=\ln{2}/2\pi$.}
		\label{Fig:constr_destr_blokh}
	\end{figure}

	The \textit{single}-passage evolution is described by the Hermitian matrix
	\begin{equation}
		\widetilde{N}=U_{\mathrm{ad}}(\tau_{\mathrm{f}},0)NU_{\mathrm{ad}}(0,\tau_\mathrm{i}).
	\end{equation}
Taking for simplicity $\tau_{\mathrm{f}}=\tau_\mathrm{a}$ and $\tau_{\mathrm{i}}=-\tau_\mathrm{a}$, we obtain the components of this matrix
	\begin{equation}
		\begin{pmatrix}
			\widetilde{N}_{11} & \widetilde{N}_{12} \\ 
			\widetilde{N}_{21} & \widetilde{N}_{22}
		\end{pmatrix}=
		\begin{pmatrix}
			\sqrt{R} & \sqrt{T}e^{2i\zeta (\tau _\mathrm{a})} \\ 
			-\sqrt{T}^{\ast }e^{-2i\zeta (\tau _\mathrm{a})} & \sqrt{R}
		\end{pmatrix}.
		\label{FullTransitionMatrix}
	\end{equation}
For a more general case, $\tau_{\mathrm{f}} \neq -\tau_{\mathrm{i}}$, see Appendix~\ref{AppendixA}.
		The wave function after the transition can be derived from Eq.~(\ref{AIMevolution}). Then the first component of the spinor becomes
	\begin{equation}
    \alpha_\mathrm{f}=\alpha_\mathrm{i}\widetilde{N}_{11}+\sqrt{1+\alpha_\mathrm{i}^2}\;e^{i\phi_\mathrm{i}}\widetilde{N}_{12}.
	\end{equation}
 	\begin{figure*}[t]
		\includegraphics[width=1.77\columnwidth]{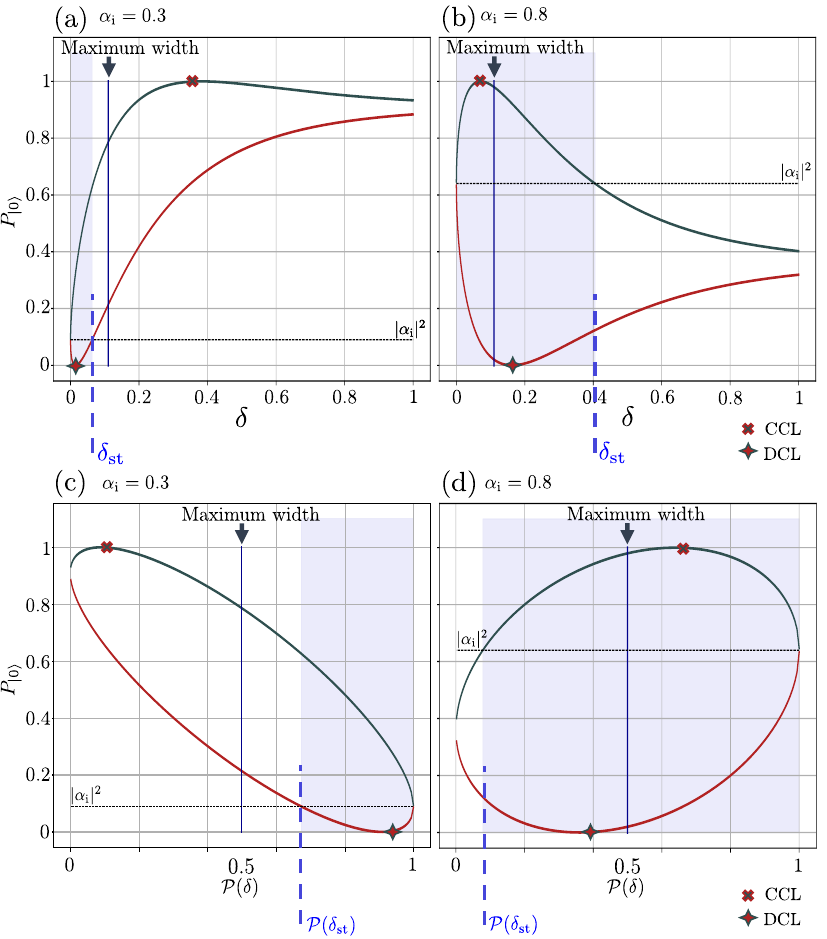}
		\caption{Visualization of the difference (or width) between constructive and destructive interference in the dependence of the occupation probability $P_{\left\vert 0\right\rangle}$ on the adiabaticity parameter $\delta$ [upper panels] and on the single-passage probability $\mathcal{P}(\delta)$ [lower panels]. (a) and (c) correspond to the case when $\alpha_\mathrm{i}=0.3$, while (b) and (d) are for $\alpha_\mathrm{i}=0.8$. The dark green curve shows the constructive case, and the dark red curve shows the destructive interference case. The light blue background region shows the values of the adiabaticity parameter where returning to the initial occupation probability is possible, that is when $\delta < \delta_{\mathrm{st}}$. The points that correspond to the long-time complete localization of the initial state are indicated: The destructive complete localization (DCL) are show with an X, and, constructive complete localization (CCL) by a star shaped like a +. The values of the adiabaticity parameter in these cases correspond to Eqs.~(\ref{DCL}, \ref{CCL}) respectively. }
		\label{Fig:Width}
	\end{figure*}
	The final occupation probability of  the $\left\vert 0\right\rangle$ state is $P_{\left\vert 0 \right\rangle \mathrm{f}}=|\alpha_\mathrm{f}|^2$. Then using Eqs.~(\ref{Zeta},\ref{PhiStokes}), we can write the direct dependence of the occupation probability on the system parameters
	\begin{equation}
		P_{\left\vert 0 \right\rangle \mathrm{f}}=\alpha_\mathrm{i}^2 e^{-2\pi\delta}+(1-\alpha_{\mathrm{i}}^2)(1-e^{-2\pi\delta})
		\nonumber
	\end{equation}
	\begin{equation}
		+2\alpha_{\mathrm{i}}\sqrt{1-\alpha_{\mathrm{i}}^2}\,e^{-\pi\delta}\sqrt{1-e^{-2\pi\delta}}\cos{\theta},
		\label{FinalProbability}
	\end{equation}
	where
	\begin{equation}
		\theta\left(\delta, \tau_\mathrm{a},\phi_\mathrm{i}\right)=\frac{\pi}{4}+\mathrm{Arg}\left [\Gamma(1-i\delta)\right]+\tau_\mathrm{a}^2+2\delta\ln{(\sqrt{2}\tau_\mathrm{a})}+\phi_\mathrm{i}.
        \label{theta}
	\end{equation}
	It is important to note that the final occupation probability does not depend on the final time; for a more general treatment see Appendix~\ref{AppendixA}. The final result will be the same as Eq.~(\ref{FinalProbability}), with $-\tau_\mathrm{a}\to\tau_\mathrm{i}$.
 
	\section{How the initial parameters affect the final probabilities}
         \label{TWOPPVOP}
	The final probability $P_{\left\vert 0 \right \rangle}$ depends on the following parameters: $\delta$, $\alpha_{\mathrm{i}}$, $\phi_{\mathrm{i}}$, $\tau_\mathrm{i}$. We are now interested in studying the contribution of the third term in Eq.~(\ref{FinalProbability}). This term is the result of interference, and (for convenience) we will designate it as $\alpha_{\mathrm{int}}^2$
	\begin{equation}
		  \alpha_{\mathrm{int}}^2= 2\alpha_{\mathrm{i}}\sqrt{1-\alpha_{\mathrm{i}}^2}\, e^{-\pi\delta}\sqrt{1-e^{-2\pi\delta}}\cos{\theta}.
            \label{AlphaInt}
	\end{equation}
	Figure~\ref{Fig:constr_destr_blokh} shows the contribution of this interference term. If the initial probability $|\alpha_\mathrm{i}|^2$ is fixed and the phase difference $\phi_\mathrm{i}$ between the components of the wave function is varied, we can obtain the final probabilities $|\alpha_\mathrm{f}|^2$ in a wide range. 
	
	Having obtained interference term we analyze the dependence on $\alpha_\mathrm{i}$ is proportional to the factor $2\alpha_{\mathrm{i}}\sqrt{1-\alpha_{\mathrm{i}}^2}$, so there is a maximum at $\alpha_\mathrm{i}=1/\sqrt{2}$. 
 
 The dependence on $\tau_\mathrm{i}$ and $\phi_\mathrm{i}$ is via the argument of $\cos{\theta}$. Since the parameters $\phi_\mathrm{i}$ and $\tau_\mathrm{i}$ both contribute to the final result only in the argument of $\mathrm{cos}$  $\theta$ in Eq.~(\ref{FinalProbability}), then their contribution is similar. 
 
 The phase before the transition is the sum of the initial phase $\phi_\mathrm{i}$ and the phase which was collected during the adiabatic evolution before the transition. The latter phase is defined with the initial time. There are minimum and maximum values of this interference term, which correspond to $\cos{\theta}=\pm 1$. We describe occupation probabilities in the diabatic basis. In our problem, we consider the system without relaxations, which means that these two levels are equivalent. Accordingly, we use the terms constructive/destructive for the final occupation probability $P_{\left\vert 0 \right\rangle}$. We define the maximum value of the occupation probability of the state $\left\vert 0 \right\rangle$ as the constructive interference case. The minimum value corresponds to the destructive interference case. These respective cases will occur if the initial phase difference is
	\begin{equation}
        \phi_\mathrm{i}^{\mathrm{destr}}=\phi_{\mathrm{i0}}+\frac{\pi}{2},
        \label{DestructivePhase}
	\end{equation}
        and
 	\begin{equation}
		\phi_\mathrm{i}^{\mathrm{constr}}=\phi_{\mathrm{i0}}-\frac{\pi}{2},
        \label{ConstructivePhase}
	\end{equation}
	where $\phi_{\mathrm{i0}}$ corresponds to the zero contribution from the interference and equals to 
	\begin{equation}
		\phi_\mathrm{i0}=2\pi n+\frac{\pi}{4}-\mathrm{Arg}\left[\Gamma(1-i\delta)\right]-\tau_\mathrm{i}^2-2\delta\ln(\sqrt{2}\tau_\mathrm{i}),
	\end{equation}
	with $n$ being an integer.         
 
	We can obtain any value of the interference term between the constructive- and destructive-interference values. The minimum and maximum values of the occupation probability are
	\begin{equation}
		P_{\left\vert 0 \right\rangle\mathrm{max}/ \mathrm{min}}=\left (\alpha_{\mathrm{i}}e^{-\pi\delta}\pm\sqrt{1-\alpha_{\mathrm{i}}^2}\sqrt{1-e^{-2\pi\delta}}\right )^2.
        \label{Probability_min_max}
	\end{equation}
	The width of this region (between destructive and constructive interference) is 
	\begin{equation}
		\Delta P (\alpha_{\mathrm{i}}, \delta)= 4\alpha_\mathrm{i}\sqrt{1-\alpha_\mathrm{i}^2}\,e^{-\pi\delta}\sqrt{1-e^{-2\pi\delta}}.
	\end{equation}
	The maximum value of the width, at $\alpha_\mathrm{i}=\frac{1}{\sqrt{2}}$, is
	\begin{equation}
		\Delta P (\delta)_{\mathrm{max}}= 2 e^{-\pi\delta}\sqrt{1-e^{-2\pi\delta}}.
	\end{equation}
	\begin{figure*}[t]
		\includegraphics[width=2\columnwidth]{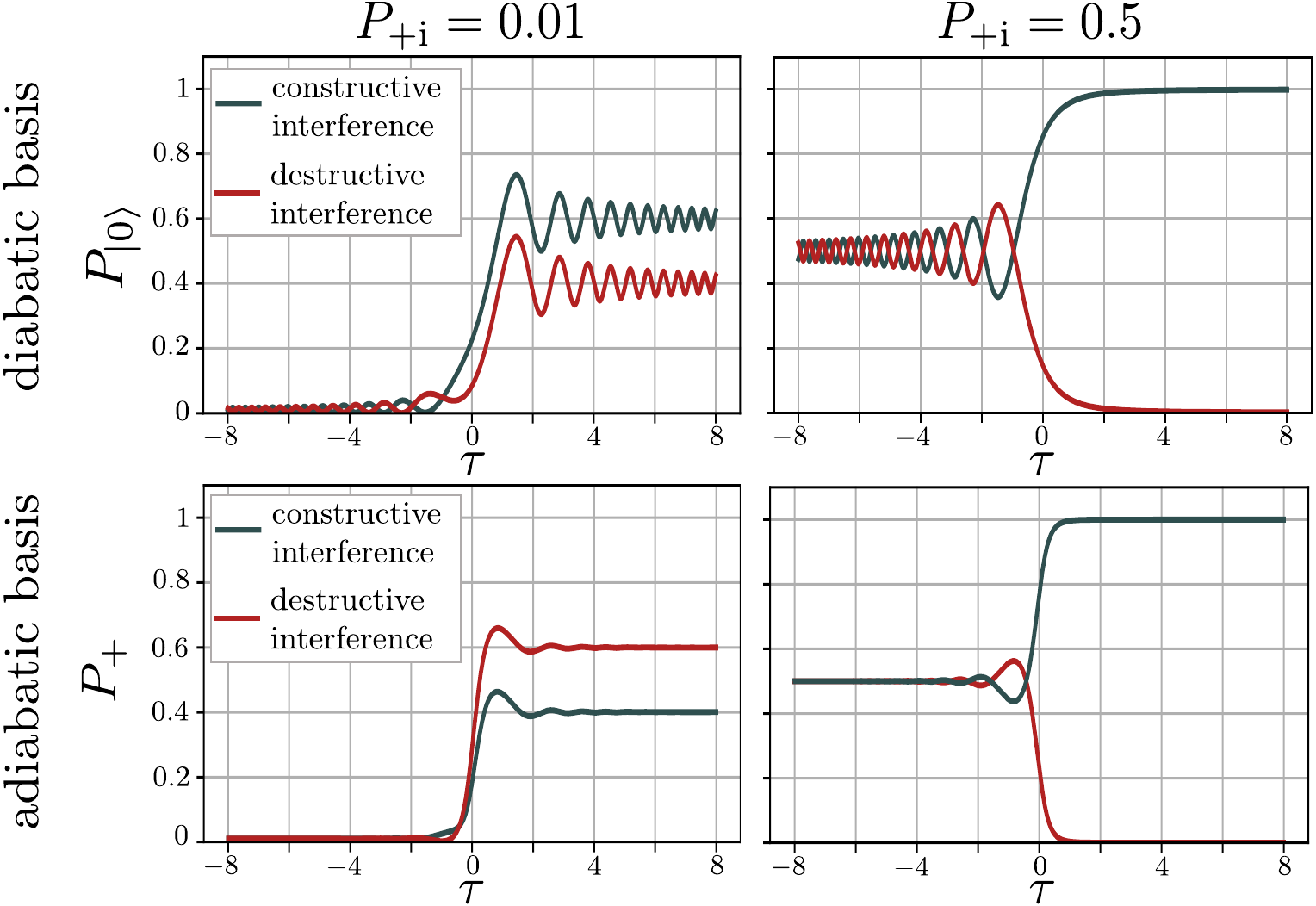}
		\caption{Multiple-passage dynamics. The plots in the first row show the dynamics for \textit{two} passages on a \textit{dia}batic basis. The left column plots show the dynamics of the \textit{single} LZSM transition with $\delta=\frac{\ln \sqrt{2}}{\pi}$ starting with an initial occupation probability $P_{\left\vert 0 \right\rangle \mathrm{i}}=0.01$. The final states are between the dark green curve (constructive interference) and the dark red curve (destructive interference). The right column shows the \textit{second} passage with the same adiabaticity parameter $\delta$ as in the first transition starting from one of the possible final stages after one transition. After the second transition, the final occupation probability could be chosen between 0 and 1 using as a tunable parameter the phase of the wave function before the second transition. The same dynamics is shown in the second row.  }
		\label{Fig:TwoPassage}
	\end{figure*}
 
	Alternatively, the width could be changed by the adiabaticity parameter $\delta$. In particular, if $\delta=\ln\sqrt{2}/\pi$, then the width becomes
	\begin{equation}
		\Delta P (\alpha_{\mathrm{i}})_{\mathrm{max}}=2\alpha_\mathrm{i}\sqrt{1-\alpha_\mathrm{i}^2}.
	\end{equation}
	If we take 
        \begin{equation}
            \alpha_\mathrm{i}=\frac{1}{\sqrt{2}} \text{ }\text{ }\mathrm{and} \text{ }\text{ }\delta=\frac{\ln\sqrt{2}}{\pi}
        \end{equation}
        at the same time, the width will be $1$, which means that the any final probability could be obtained.

        Figure~\ref{Fig:Width} shows the dependence of the width between constructive and destructive interference on the adiabaticity parameter $\delta$ as well as on the single-passage probability $\mathcal{P}(\delta)$ in two cases of the initial occupation probability. The dark green line shows the maximal value of Eq.~(\ref{AlphaInt}) and the dark red line shows the minimal value. The impact of the interference term $\alpha_\mathrm{int}^2$ depends on the adiabaticity parameter. When $\delta\gg 1$ or $\delta\ll 1$, the impact of the interference term tends to zero. Values of the adiabaticity parameter, when the interference impact is not negligible, correspond to the fast processes which are not adiabatic.

	\section{Condition for returning to the same occupation probability as the initial one after the transition}
 
         \label{RITISATT}
		The condition for returning to the same state as the initial one after the transition could be found from Eq.~(\ref{FinalProbability}) if we require $P_{\left\vert 0 \right\rangle}=\alpha_{\mathrm{i}}^2$, which determines the value of $\cos \theta$:
	\begin{equation}
		\cos{\theta}=\frac{\left(2\alpha_{\mathrm{i}}^2-1\right)\sqrt{1-\exp(-2\pi\delta)}}{2\alpha_{\mathrm{i}}\sqrt{1-\alpha_{\mathrm{i}}^2}\exp(-\pi\delta)}.
	\end{equation}
        This condition can be satisfied when the value of $\left\vert\cos \theta\right\vert\leq 1$, which requires
	\begin{equation}
		\delta\leq\delta_\mathrm{st}\equiv\frac{1}{2\pi}\ln{\frac{1}{\left(2\alpha_{\mathrm{i}}^2-1\right)}} =-\frac{1}{2\pi}\ln{\left(2\alpha_{\mathrm{i}}^2-1\right)}.
        \label{DeltaST}
	\end{equation}
	Then the phase difference between $\alpha_{\mathrm{i}}$ and $\beta_{\mathrm{i}}$ should be
	\begin{equation}
		\phi_{\mathrm{i}}=\arccos{\frac{\left(2\alpha_{\mathrm{i}}^2-1\right)\sqrt{1-e^{-2\pi\delta}}}{2\alpha_{\mathrm{i}}\sqrt{1-\alpha_{\mathrm{i}}^2}\;e^{-\pi\delta}}}+\phi_\mathrm{i0}-\frac{\pi}{2}.
        \label{PhaseStable}
	\end{equation}

    In Fig.~\ref{Fig:Width} the region where there could be a return to the initial occupation probability is shown as a light blue region. In the \textit{diabatic} basis, it corresponds to the small values of the adiabaticity parameter $\delta$. In these cases the processes are nonadiabatic.   

     Consider now reaching the ground and excited states, which correspond to CLs. In Fig.~\ref{Fig:Width} we indicate the points where it is possible to obtain the system in the ground state. If the final occupation probability is equal to one this means that the final state is $\left\vert 0\right\rangle$. In the opposite case, if the final occupation probability is equal to zero, the state is $\left\vert 1\right\rangle$. 
    
    These cases are present for any initial occupation probability and correspond to the minimum and maximum of the destructive and constructive interference occupation probabilities. From Eq.~(\ref{Probability_min_max}) the values of the adiabaticity parameter $\delta$ for these cases could be found. For destructive CL, the adiabaticity parameter is  
    \begin{equation}
        \delta_{\mathrm{DCL}}=-\frac{1}{2\pi}\ln{\left(1-\alpha_{\mathrm{i}}^2\right)},
        \label{DCL}
    \end{equation}
    and for constructive CL, the adiabaticity parameter becomes
    \begin{equation}
        \delta_{\mathrm{CCL}}=-\frac{1}{\pi}\ln{\alpha_\mathrm{i}}.
        \label{CCL}
    \end{equation}

The results of this section might appear to be surprising because these indicate that \textit{it is possible for the occupation probability to return to its initial value by only using phase control}. For any initial occupation probability, it is possible to steer the state to a desired final long-time target state.

	\section{Preparing a desired target state by only changing its initial phase}
 	
         \label{PTSS}
	We now consider \textit{how to control qubits by changing their initial phase}. We describe a qubit undergoing adiabatic evolution under the constant offset 
    \begin{equation}
        \varepsilon (t)=\mathrm{const}=\varepsilon_0
    \end{equation}
in Eq.~(\ref{Schrodingerequation}). If we apply a constant drive, the evolution of the wave function consists in changing the phase. In the Bloch sphere, this corresponds to a rotation around the $z$ axis. The phase which appears after this evolution during the time $t_\mathrm{wait}$ is
	\begin{equation}
		\phi_{\mathrm{ad}}=-\mathrm{sgn}(\varepsilon_0)\frac{\sqrt{\Delta^2+\varepsilon_0^2}}{\hbar}\,t_\mathrm{wait}.
        \label{Eq.daiabatic_phase}
	\end{equation}
    The sign of the phase depends on the sign of the drive $\varepsilon_0$, see details in Appendix~\ref{AppendixB}. This means that using this type of signal we can change the phase of the wave function without changing the occupation probability. We can use it to control the phase to prepare the desired initial state before the LZSM transition. 
 
    As was discussed in the previous section, for any initial occupation probability $P_{\left\vert 0\right\rangle}$, it is possible to obtain the ground state. This means that by changing the adiabaticity parameter $\delta$ and the initial phase $\phi_\mathrm{i}$ it is possible to obtain any desired final state. 

   If the initial state is close to the ground state, an infinite adiabaticity parameter $\delta$ is needed to obtain another ground state. To overcome this difficulty, we can apply multiple passages. Thus, two or more LZSM transitions allow to obtain the final occupation probability with experimentally realizable parameters.
    
    For illustration, Figure~\ref{Fig:TwoPassage} demonstrates two passages (in the diabatic basis see the 
 panels in the first row). The first passage starts with an initial occupation probability $P_{\left\vert 0 \right\rangle\mathrm{i}}=0.01$. This state is close to the ground state $\left\vert 1\right\rangle$. After the first LZSM transition (left panel) with the adiabaticity parameter $\delta=\ln \sqrt{2}/\pi$, the final occupation probability could be $P_{\left\vert 0 \right\rangle\mathrm{f}}\approx 0.5\pm 0.1$. Applying the second LZSM transition with the same adiabaticity parameter as one of the possible final states after the first transition, the final occupation probability is now in the range between 0 and 1. Thus, changing the phase of the wave function after the first transition gives the possibility to choose the final state after the next transition.  

    This process can be also considered on the adiabatic basis which is the eigen energy levels basis. All these formulas could be obtained by the relation between diabatic and adiabatic basis; see details in Appendix~\ref{AppendixC}. The example of the double passage in the adiabatic basis is shown in the second row in Fig.~\ref{Fig:TwoPassage}.

    After the single LZSM passage, any final state could be obtained using as the tunable parameters the initial phase and adiabaticity parameter $\delta$. The multi-passage dynamics with finite adiabaticity parameter could be applied in cases when obtaining some final stages of the infinite adiabaticity parameter is required for the single-passage case.

	\section{Conclusions}
 \begin{table*}[ht]
		\begin{center}
			\begin{tabularx}{18.4775cm}{|C{6cm}|C{6cm}|C{6cm}|}
				\hline
				\rule{0pt}{3ex} & \textbf{Initial phase $\varphi_\mathrm{i}$} & \\
                \rule{0pt}{3ex} \textbf{ Type of dynamics} & \textbf{and} &  \textbf{Final occupation probability}   \\
                 & \textbf{adiabaticity parameter $\delta$} & \textbf{$P_{\left\vert 0\right\rangle\mathrm{f}}$} \\[2ex]
    \hline
				 \rule{0pt}{4ex} Zero interference case& $\phi_{\mathrm{i}0}=2\pi n+\frac{\pi}{4}-\mathrm{Arg}\left[\Gamma(1-i\delta)\right]$ & $\alpha_\mathrm{i}^2 e^{-2\pi\delta}+(1-\alpha_{\mathrm{i}}^2)(1-e^{-2\pi\delta})$
				\\
                    \rule{0pt}{3ex}&$-\tau_\mathrm{i}^2-2\delta\ln(\sqrt{2}\tau_\mathrm{i})$& \\[2ex]  \hline
                    \rule{0pt}{4ex} Destructive interference
& $\phi_\mathrm{i}^{\mathrm{destr}}=\phi_{\mathrm{i0}}+\frac{\pi}{2}$ & $\left (\alpha_{\mathrm{i}}e^{-\pi\delta}-\sqrt{1-\alpha_{\mathrm{i}}^2}\sqrt{1-e^{-2\pi\delta}}\right )^2$
				\\[2ex]  \hline
                \rule{0pt}{4ex} Constructive interference
& $\phi_\mathrm{i}^{\mathrm{constr}}=\phi_{\mathrm{i0}}-\frac{\pi}{2}$ & $\left (\alpha_{\mathrm{i}}e^{-\pi\delta}+\sqrt{1-\alpha_{\mathrm{i}}^2}\sqrt{1-e^{-2\pi\delta}}\right)^2$ 
                    \\[2ex]  \hline
                 \rule{0pt}{4ex} 
& $\phi_{\mathrm{i}}=\arccos{\frac{\left(2\alpha_{\mathrm{i}}^2-1\right)\sqrt{1-e^{-2\pi\delta}}}{2\alpha_{\mathrm{i}}\sqrt{1-\alpha_{\mathrm{i}}^2}e^{-\pi\delta}}}+\phi_\mathrm{i0}-\frac{\pi}{2}$ & 	\\ [0ex]
    \rule{0pt}{2ex} Returning to initial occupation probability & and & Initial occupation probability $P_{\left\vert 0\right\rangle \mathrm{i}}$\\ [2ex]
    \rule{0pt}{2ex} & $\delta\leq-\frac{1}{2\pi}\ln{\left(2\alpha_{\mathrm{i}}^2-1\right)}$ & \\[2ex]
    \hline        
    \rule{0pt}{4ex} & $\phi_\mathrm{i}=\phi_\mathrm{i}^{\mathrm{destr}}$ & \\[2ex]
    \rule{0pt}{2ex} Destructive complete localization in a target state
 & and & $0$ \\[2ex]
    \rule{0pt}{2ex} & $\delta_{\mathrm{DCL}}=-\frac{1}{2\pi}\ln{\left(1-\alpha_{\mathrm{i}}^2\right)}$ & \\[2ex]
    \hline
    \rule{0pt}{4ex} & $\phi_\mathrm{i}=\phi_\mathrm{i}^{\mathrm{constr}}$ & \\[2ex]
    \rule{0pt}{2ex} Constructive complete localization in a target state
 & and & $1$ \\[2ex]
    \rule{0pt}{2ex} & $\delta_{\mathrm{CCL}}=-\frac{1}{\pi}\ln{\alpha_{\mathrm{i}}}$ & \\[2ex]
    \hline
			\end{tabularx}
		\end{center}
		\caption{Summary of the main results. If the adiabaticity parameter $\delta$ can have any value, we do not show it in the second column, like in the first three lines. }
		\label{Table:Conclusions}
	\end{table*}  
 
	We analyzed the single-passage qubit dynamics with a linear drive $\varepsilon(t)=vt$. The initial phase of the wave function was considered as a parameter. We derived several results, including how to recover the initial occupation probability (analog of transitionless driving), as well as steering the final state into either the ground or the excited states (CL). Results are summarized in Table I. We obtained various target occupation probabilities using phase control. Also, multiple passages were considered in order to obtain desired qubit states for a given value of the adiabaticity parameter $\delta$. Results are shown in Table~\ref{Table:Conclusions}.


	\begin{acknowledgments}
 
		The authors acknowledge fruitful discussions with Oleg Ivakhnenko.
		The research work of P.O.K. and S.N.S. is sponsored by the Army Research Office under Grant No.~W911NF-20-1-0261. P.O.K. gratefully acknowledges an IPA RIKEN scholarship. F.N. is supported in part by:
Nippon Telegraph and Telephone Corporation (NTT) Research, the Japan Science and Technology Agency (JST) [via the Quantum Leap Flagship Program (Q-LEAP), and the Moonshot R\&D Grant Number JPMJMS2061], the Asian Office of Aerospace Research and Development (AOARD) (via Grant No. FA2386-20-1-4069), and the Office of Naval Research (ONR) Grant No.N62909-23-1-2074.
	\end{acknowledgments}
       
        \newpage

	\appendix
	
	\section{Different initial and final times}
	\label{AppendixA}
	The values of the initial and final times are not the same in the general case. In this section, we will show the generalization of the transfer matrix, Eq.~(\ref{FullTransitionMatrix}). Different initial and final times correspond to the value of phase which will occur due to adiabatic evolution. We use the following notations for values of the initial and final times respectively $\tau_{\mathrm{i}}$ and $\tau_{\mathrm{f}}$. Following Ref.~\cite{Kofman2023} the adiabatic evolution matrix before the transition is
	\begin{equation}
		\tau <0:\,\,\,\,U_{\mathrm{ad}}(0, \tau_\mathrm{i})=\left( 
		\begin{array}{cc}
			\exp \left( -i\zeta(\tau_{\mathrm{i}}) \right) & 0 \\ 
			0 & \exp \left( i\zeta(\tau_{\mathrm{i}}) \right)%
		\end{array}%
		\right)
	\end{equation}%
	and after the transition
	\begin{equation}
		\tau >0:\,\,\,\,U_{\mathrm{ad}}( \tau_\mathrm{f}, 0)=\left( 
		\begin{array}{cc}
			\exp \left( i\zeta(\tau_\mathrm{f}) \right) & 0 \\ 
			0 & \exp \left( -i\zeta(\tau_\mathrm{f}) \right)%
		\end{array}%
		\right) .
	\end{equation}
	Now we consider a bias which is linear in time $\varepsilon (t)=vt$. The asymptotic expressions for $\zeta $ at large times, i.e. at $t=\pm \tau _{\mathrm{a}}\sqrt{2\hbar /v}$, with $\tau _{\mathrm{a}}\gg1$ defined in Eq.~(\ref{Zeta}).

	In general, Eq.~(\ref{FullTransitionMatrix}) becomes
	\widetext
	\begin{equation}
		\widetilde{N}=\begin{pmatrix}
			\exp \left(i\zeta(\tau_{\mathrm{f}})-i \zeta(\tau_{\mathrm{i}})\right)\sqrt{\mathcal{P}} & \exp \left(i\zeta(\tau_{\mathrm{f}})+i \zeta(\tau_{\mathrm{i}})+i2\varphi_{\mathrm{S}}\right)\sqrt{1-\mathcal{P}}\\
			-\exp \left(-i\zeta(\tau_{\mathrm{f}})-i \zeta(\tau_{\mathrm{i}})-i2\varphi_{\mathrm{S}}\right)\sqrt{1-\mathcal{P}} & \exp \left(-i\zeta(\tau_{\mathrm{f}})+i \zeta(\tau_{\mathrm{i}})\right)\sqrt{\mathcal{P}}
		\end{pmatrix}.
	\end{equation}
	The final components of the wave function then become 
	\begin{equation}
		\alpha_{\mathrm{f}}=\exp \left[i\zeta(\tau_{\mathrm{f}})-i \zeta(\tau_{\mathrm{i}})\right]\sqrt{\mathcal{P}}\alpha_{\mathrm{i}}+\exp \left[i\zeta(\tau_{\mathrm{f}})+i \zeta(\tau_{\mathrm{i}})+i2\varphi_{\mathrm{S}}+i\phi_{\mathrm{i}}\right]\sqrt{1-\mathcal{P}}\sqrt{1-\alpha_{\mathrm{i}}^2},
	\end{equation}
	\begin{equation}
		\beta_{\mathrm{f}}=-\exp \left[-i\zeta(\tau_{\mathrm{f}})-i \zeta(\tau_{\mathrm{i}})-i2\varphi_{\mathrm{S}}\right]\sqrt{1-\mathcal{P}}\alpha_{\mathrm{i}}+\exp \left[-i\zeta(\tau_{\mathrm{f}})+i \zeta(\tau_{\mathrm{i}})+i\phi_{\mathrm{i}}\right]\sqrt{\mathcal{P}}\sqrt{1-\alpha_{\mathrm{i}}^2}.
	\end{equation}
	\endwidetext
	This gives the final state depending on the initial phase difference between the spinor components. Knowing this dependence is important when we want to describe the dynamics with multiple passages. If we are interested in the occupation probability, the final time does not matter when it is large enough for applying the adiabatic-impulse model. It is understandable that after the transition (when the energy levels are far from each other) the final time enters only at the phase that was collected during the adiabatic evolution.
	
	\section{Adiabatic evolution in the diabatic basis}
	\label{AppendixB}
 	\begin{figure}[t]
		\includegraphics[width=1\columnwidth]{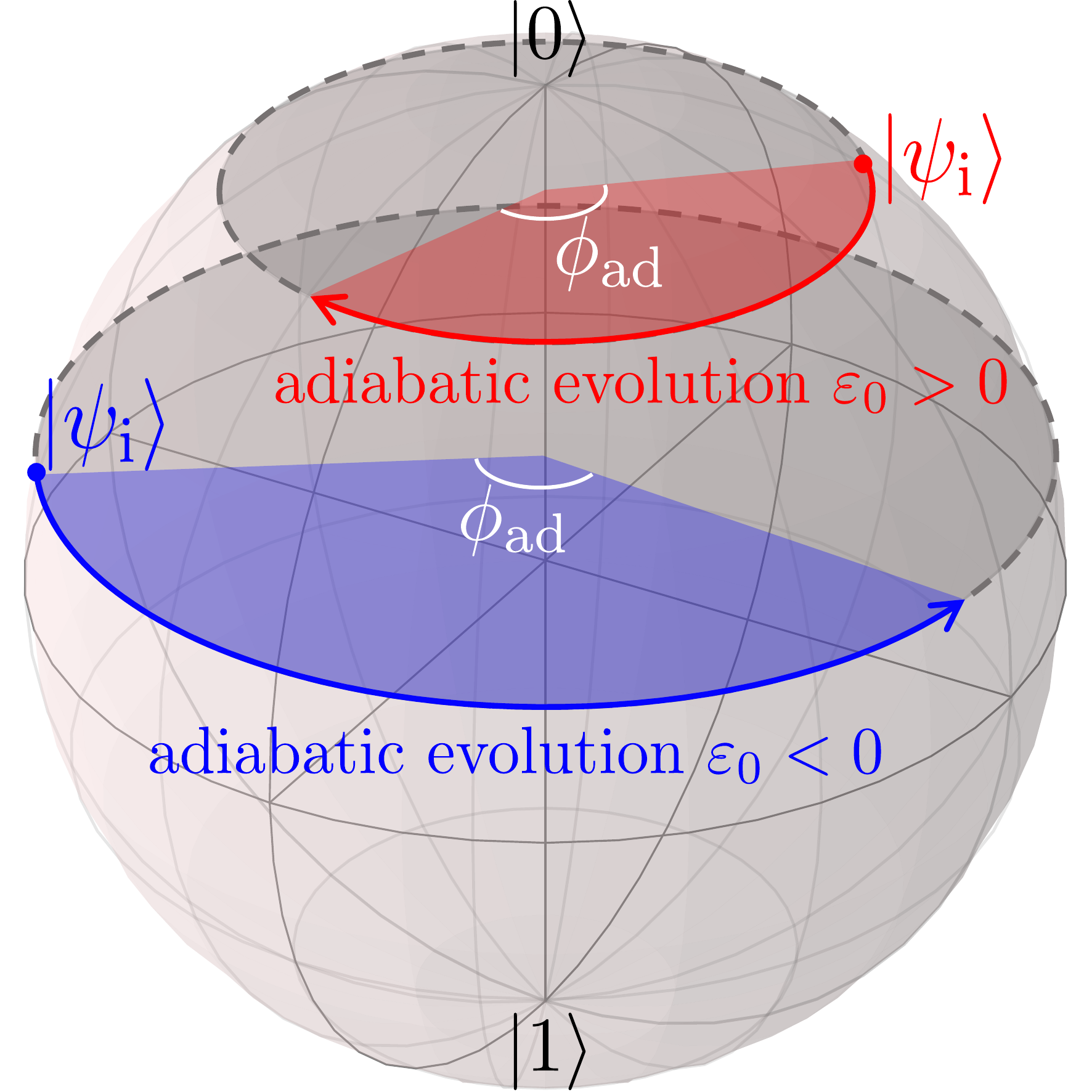}
		\caption{Adiabatic evolution on the Bloch sphere. The red curve shows that when the drive is constant and $\varepsilon_0 > 0$, the evolution corresponds to a clockwise (seeing from the top) rotation around the \textit{z}-axis accumulating the phase $\phi_\mathrm{ad}$, which is defined in Eq.~(\ref{Eq.daiabatic_phase}). If the bias is $\varepsilon_0 < 0$, the rotation turns counterclockwise, shown in blue.}
		\label{Fig:BlochSphere}
	\end{figure}
	If we want to prepare the state with some special phase we can use the adiabatic evolution. This evolution is described by the following matrix in the adiabatic basis 
	\begin{equation}
		U_{\mathrm{ad}}(t_\mathrm{f}, t_\mathrm{i})=\left( 
		\begin{array}{cc}
			\exp \left( -i\zeta \right) & 0 \\ 
			0 & \exp \left( i\zeta \right)
		\end{array}
		\right),
	\end{equation}
	where
	\begin{eqnarray}
		\zeta \left(t_\mathrm{f}, t_\mathrm{i}\right) &=&\frac{1}{2\hbar }%
		\int\limits_{t_\mathrm{i}}^{t_\mathrm{f}}\sqrt{\Delta ^{2}+\varepsilon(t)^2}%
		d t.
        \label{zeta}
	\end{eqnarray}
	If the drive is constant $\varepsilon(t)=\varepsilon_0$, the evolution is described by
	\begin{equation}
		U_{\mathrm{ad}}(t_\mathrm{f}, t_\mathrm{i})=\left( 
		\begin{array}{cc}
			\exp \left[ -i\omega t_\mathrm{wait} \right] & 0 \\ 
			0 & \exp \left[ i\omega t_\mathrm{wait} \right]
		\end{array}
		\right),
	\end{equation}
	where we introduce the following notation $t_{\mathrm{wait}}=t_{\mathrm{f}}-t_{\mathrm{i}}$ and
	\begin{equation}
		\omega=\frac{\sqrt{\Delta^2+\varepsilon_0^2}}{2\hbar}.
	\end{equation}
	The adiabatic and diabatic bases are related in the following way
	\begin{equation}
		\left\vert\varphi_{\pm}\right\rangle=\gamma_{\mp}\left\vert\psi_-\right\rangle\mp\gamma_{\pm}\left\vert\psi_+\right\rangle,
	\end{equation}
	where
	\begin{equation}
		\gamma_{\pm}=\frac{1}{\sqrt{2}}\sqrt{1\pm\frac{\varepsilon(t)}{\sqrt{\Delta^2+\varepsilon(t)^2}}}.
	\end{equation}
	The matrix which describes the transition from the diabatic basis to the adiabatic one is
	\begin{equation}
		\mathcal{M}=\begin{pmatrix}
			\gamma_- & -\gamma_+\\
			\gamma_+ & \gamma_-
            \label{matrixDiabaticToDiabatic}
		\end{pmatrix}.
	\end{equation}
	The matrix $\mathcal{M}$ is unitary, so the adiabatic evolution in the diabatic basis is described by the matrix
	\begin{equation}
		U_{\mathrm{ad}}^{\mathrm{diab}}(t_\mathrm{f},t_\mathrm{i})=\mathcal{M}^T U_{\mathrm{ad}}(t_\mathrm{f},t_\mathrm{i})\mathcal{M},
	\end{equation}
	\widetext
	\begin{equation}
		U_{\mathrm{ad}}^{\mathrm{diab}}(t_\mathrm{f},t_\mathrm{i})=\begin{pmatrix}
			\gamma_-^2 e^{-i\omega t_{\mathrm{wait}}}+ \gamma_+^2 e^{i\omega t_{\mathrm{wait}}} & \gamma_-\gamma_+\left(\exp \left[ i\omega t_{\mathrm{wait}}\right]-\exp \left[ -i\omega t_{\mathrm{wait}} \right]\right) \\
            \\
			\gamma_-\gamma_+\left(\exp \left[ i\omega t_{\mathrm{wait}} \right]-\exp \left[ -i\omega t_{\mathrm{wait}} \right]\right) & \gamma_+^2 \exp \left[ -i\omega t_{\mathrm{wait}} \right]+ \gamma_-^2 \exp \left[ i\omega t_{\mathrm{wait}} \right]
		\end{pmatrix}.
	\end{equation}
	\endwidetext \!\!\!\!\!\!The same result could be obtained by directly solving the Schr\"{o}dinger equation in the diabatic basis.
	
	If the adiabatic evolution is far from the crossing energy levels region, $|\varepsilon_0|\gg\Delta$, the adiabatic evolution will be
	\begin{equation}
		\varepsilon_0 <0:\,\,U_{\mathrm{ad}}^{\mathrm{diab}}(t_\mathrm{f},t_\mathrm{i})=\left( 
		\begin{array}{cc}
			\exp \left[ -i\omega t_{\mathrm{wait}} \right] & 0 \\ 
			0 & \exp \left[ i\omega t_{\mathrm{wait}} \right]%
		\end{array}%
		\right);
	\end{equation}
	\begin{equation}
		\varepsilon_0 >0:\,\,U_{\mathrm{ad}}^{\mathrm{diab}}(t_\mathrm{f},t_\mathrm{i})=\left( 
		\begin{array}{cc}
			\exp \left[ i\omega t_{\mathrm{wait}} \right] & 0 \\ 
			0 & \exp \left[ -i\omega t_{\mathrm{wait}} \right]%
		\end{array}%
		\right).
	\end{equation}
	It shows that (for the different signs of $\varepsilon_0$) the rotations along the $z$ axis are in different directions, and the visualization is presented in Fig.~\ref{Fig:BlochSphere}. 
 	\begin{figure*}[t]
		\includegraphics[width=2\columnwidth]{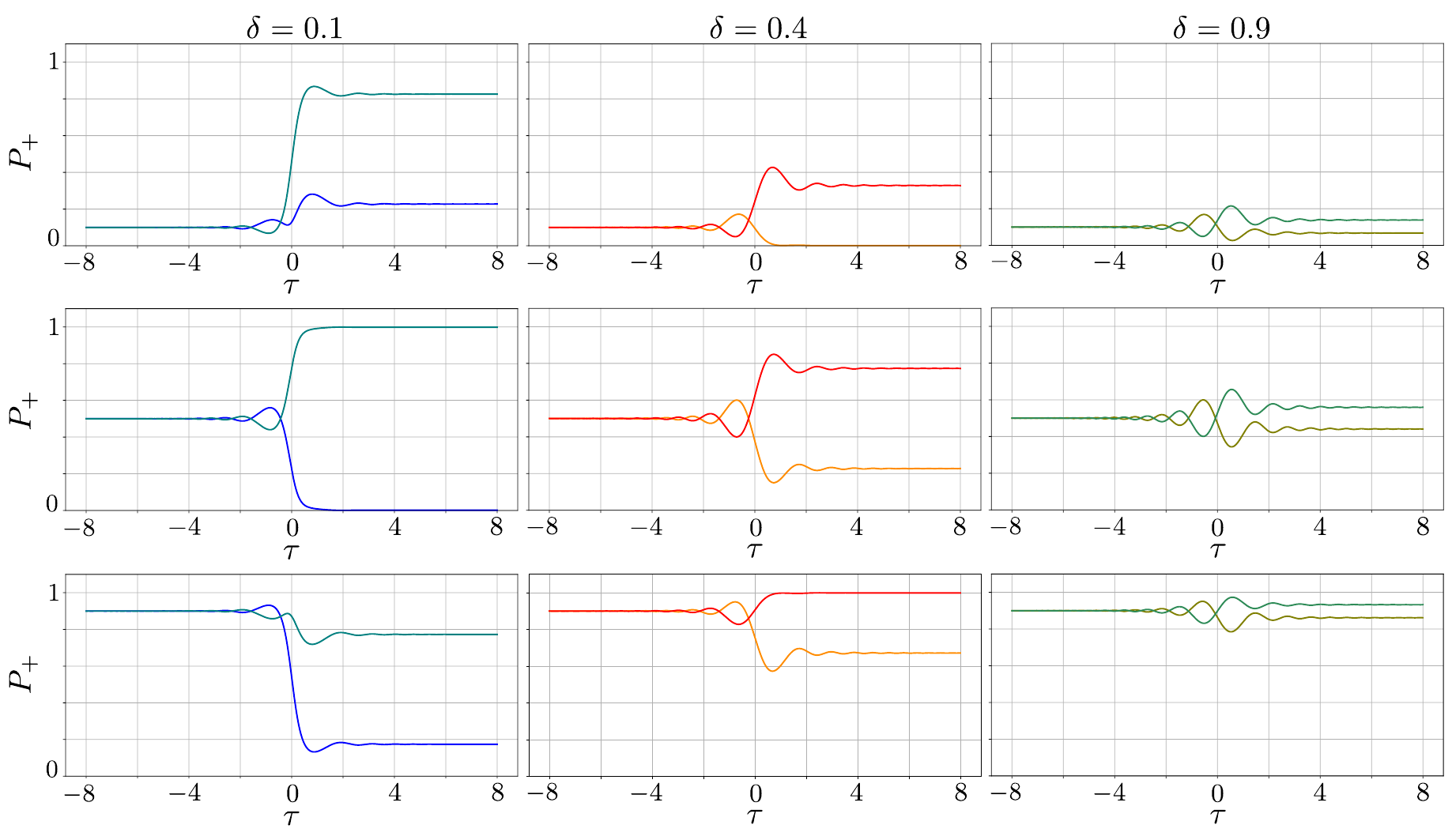}
		\caption{Dependence of the occupation probability on time in cases of constructive/destructive interference for different adiabaticity parameters $\delta$ and for different initial occupation probabilities $P_{+\mathrm{i}}$ in the adiabatic basis.  The three columns correspond to three values of the adiabaticity parameter $\delta$. Each panel shows the dynamics of the occupation probability $\mathcal{P}_+$ versus the dimensionless time $\tau$. These dynamical evolutions show that the impact of the interference is smaller when $\delta$ is larger. Maximal width is when the initial occupation probability is equal to $0.5$ (see the second row). }
		\label{Fig:MinMax}
	\end{figure*}

        \section{Adiabatic basis}
         \label{AppendixC}
	The matrix which describes the transition from a diabatic basis to the adiabatic one is Eq.~(\ref{matrixDiabaticToDiabatic}). Using this relation and assuming that we are far from the quasi-crossing region, we obtain the transition matrix in the adiabatic basis
    \begin{equation}
        \widetilde{N}^{\mathrm{ad}}=\begin{pmatrix}
            \left(\sqrt{T}\right)^* e^{-2i\zeta(\tau_{\mathrm{a}})} & -\sqrt{R}\\
            \sqrt{R} & \sqrt{T}e^{2i\zeta(\tau_{\mathrm{a}})}
        \end{pmatrix}.
    \end{equation}
    The wave function in the adiabatic basis is 
    \begin{equation}
        \left\vert\psi^\mathrm{ad}\right\rangle=\begin{pmatrix}
            b_1\\
            b_2
        \end{pmatrix}.
    \end{equation}
    Then if $b_{1\mathrm{i}}$ and $b_{2\mathrm{i}}$ are the initial components of the spinor, the final components become
    \begin{equation}
        b_{1\mathrm{f}}=\sqrt{1-\mathcal{P}}\exp{\left[-i(2\zeta(\tau_\mathrm{a}+\varphi_\mathrm{S})\right]}b_{1\mathrm{i}}-\sqrt{\mathcal{P}}b_{2\mathrm{i}},
    \end{equation}
    \begin{equation}
        b_{1\mathrm{f}}=\sqrt{\mathcal{P}}b_{1\mathrm{i}}+\sqrt{1-\mathcal{P}}\exp{\left[i(2\zeta(\tau_\mathrm{a}+\varphi_\mathrm{S})\right]}b_{2\mathrm{i}}.
    \end{equation}
    Then the occupation probability becomes
    \begin{equation}
        P_{+}=|b_{1\mathrm{f}}|^2=(1-\mathcal{P})b_{1\mathrm{i}}^2+\mathcal{P}b_{2\mathrm{i}}^2+2\sqrt{\mathcal{P}(1-\mathcal{P})}b_{1\mathrm{i}}b_{2\mathrm{i}}\cos \theta,
    \end{equation}
    where $\theta$ is defined in Eq.~(\ref{theta}) and the initial components of the spinor satisfy the following relation $b_{2\mathrm{i}}=\sqrt{1-b_{1\mathrm{i}}^2}e^{i\phi_\mathrm{i}}$.

    The condition for staying in the same state as the initial one after the transition is given by
    \begin{equation}
		\cos{\theta}=\frac{\left(2b_{1\mathrm{i}}^2-1\right)\sqrt{\mathcal{P}}}{2b_{1\mathrm{i}}\sqrt{1-b_{1\mathrm{i}}^2}\sqrt{1-\mathcal{P}}}.
    \end{equation}
    The condition of the existence of $\cos{\theta}$ becomes
    \begin{equation}
        \delta > \delta_{\mathrm{st}}^{\mathrm{ad}} \equiv -\frac{1}{2\pi}\ln \left[4b_{1\mathrm{i}}^2(1-b_{1\mathrm{i}}^2)\right].
    \end{equation}

    \section{Generalization of the problem}
    \label{AppendixD}
    If we consider a non-linear perturbation which could be described by the adiabatic-impulse model, then all the above formulas could be generalized. According to the adiabatic-impulse model, the dynamics will be separated on three stages: adiabatic
    \begin{equation}
        U_\mathrm{ad}=\begin{pmatrix}
            \exp(-i\zeta_1) & 0\\
            0 & \exp(i\zeta_1)
        \end{pmatrix};
    \end{equation}
    transition
    \begin{equation}
        N=\begin{pmatrix}
            N_{11} & N_{12}\\
            N_{21} & N_{22}
        \end{pmatrix};
    \end{equation}
    and adiabatic again
        \begin{equation}
        U_\mathrm{ad}=\begin{pmatrix}
            \exp(i\zeta_2) & 0\\
            0 & \exp(-i\zeta_2)
        \end{pmatrix},
    \end{equation}
    where $\zeta$ is defined in Eq.~(\ref{zeta}). The initial state is the same as Eq.~(\ref{InitilState}). Then the components of the wave function are  
    \begin{equation}
        \alpha_\mathrm{f}=N_{11}\alpha_\mathrm{i}\exp\left[i(\zeta_2-\zeta_1)\right]+N_{12}\beta_\mathrm{i}\exp\left[i(\zeta_1+\zeta_2)\right],
    \end{equation}
    \begin{equation}
        \beta_\mathrm{f}=N_{21}\alpha_\mathrm{i}\exp\left[-i(\zeta_1+\zeta_2)\right]+N_{22}\beta_\mathrm{i}\exp\left[i(\zeta_1-\zeta_2)\right].
    \end{equation}
    Then the occupation probability of $\left\vert0\right\rangle$ becomes
    \begin{equation}
        |\alpha_\mathrm{f}|^2=|N_{11}|^2\alpha_\mathrm{i}^2+|N_{12}|^2(1-\alpha_\mathrm{i}^2)
        \nonumber
    \end{equation}
    \begin{equation}
        +2\alpha_\mathrm{i}\sqrt{1-\alpha_\mathrm{i}^2}|N_{11}||N_{12}|\cos\left(2\zeta_1+\phi_\mathrm{i}+\varphi_{11}-\varphi_{12}\right),
    \end{equation}
    where the components of the transfer matrix are rewritten as $N_{11}=|N_{11}|e^{i\varphi_{11}}$ and $N_{12}=|N_{12}|e^{i\varphi_{12}}$.
    As a result, we see that all the dependence on phase is in the cosine at the one term which is associated with interference. It shows that all results are applicable not only for the linear perturbation but also for any perturbation which could be approximated by adiabatic-impulse model  \cite{Kofman2023}.

    In Fig.~\ref{Fig:MinMax} we show the dependence of the constructive and destructive interference versus time for different initial occupation probabilities and different adiabaticity parameters $\delta$ in the adiabatic basis.  
	\newpage
	\nocite{apsrev41Control} 
	\bibliographystyle{apsrev4-1}
	\bibliography{LZSM2,1}
	
\end{document}